\documentclass[runningheads]{llncs}
\usepackage{graphicx}
\usepackage{amsfonts}
\usepackage{calligra}
\usepackage{amsmath}
\usepackage{subfigure}
\usepackage{multirow}

\begin{document}
\title{Deep Attentive Wasserstein Generative Adversarial Networks for MRI Reconstruction with Recurrent Context-Awareness}

\titlerunning{Deep Attentive Wasserstein Generative Adversarial Networks}
%
\author{Yifeng Guo\inst{1} \and
Chengjia Wang\inst{2} \and
Heye Zhang\inst{1} \and
Guang Yang\inst{3,4}}
\authorrunning{Y. Guo et al.}
%

\institute{Sun Yat-Sen University, Guangzhou, China \and
University of Edinburgh, Edinburgh EH16 4TJ, UK \and
Cardiovascular Research Centre, Royal Brompton Hospital, London SW3 6NP, UK \and
National Heart \& Lung Institute, Imperial College London, London SW7 2AZ, UK}
\maketitle              

\begin{abstract}
The performance of traditional compressive sensing-based MRI (CS-MRI) reconstruction is affected by its slow iterative procedure and noise-induced artefacts. Although many deep learning-based CS-MRI methods have been proposed to mitigate the problems of traditional methods, they have not been able to achieve more robust results at higher acceleration factors. Most of the deep learning-based CS-MRI methods still can not fully mine the information from the \textit{k}-space, which leads to unsatisfactory results in the MRI reconstruction. In this study, we propose a new deep learning-based CS-MRI reconstruction method to fully utilise the relationship among sequential MRI slices by coupling Wasserstein Generative Adversarial Networks (WGAN) with Recurrent Neural Networks. Further development of an attentive unit enables our model to reconstruct more accurate anatomical structures for the MRI data. By experimenting on different MRI datasets, we have demonstrated that our method can not only achieve better results compared to the state-of-the-arts but can also effectively reduce residual noise generated during the reconstruction process.
\end{abstract}
\section{Introduction}
Compressed sensing magnetic resonance imaging (CS-MRI) \cite{donoho2006compressed} has been proposed for accelerating MRI process. This technique uses a small fraction of data to reconstruct images from sub-Nyquist sampling. Assuming the raw data is compressible, CS-MRI performs nonlinear optimisations on the undersampling data without sacrificing the quality of the reconstructed images significantly. 

However, it is still very challenging to consolidate the speed of reconstruction and robustness of image quality maintenance in one CS-MRI based framework. On the one hand, CS-MRI tries to solve underdetermined equations to perceive the original signals from the limited undersampled data. This requires nonlinear optimisation solvers for a common non-convex system that usually involve iterative computations, which can result in prolonged reconstruction time \cite{hollingsworth2015reducing}. On the other hand, CS-MRI may produce images with degraded image quality and low signal-to-noise ratio (SNR) from randomly highly undersampled \textit{k}-space data \cite{Ma2008An}. Moreover, in addition to a large amount of computation needed for the nonlinear optimisation, CS-MRI also requires that the acquisition matrix and the sparse transformation matrix are unrelated. Based on the above limitations, the acceleration factor of CS-MRI is generally between 2 and 6.

Recently, deep learning-based CS-MRI methods have emerged as an effective way to solve the problems of slow and unstable MRI reconstruction ~\cite{wang2016accelerating,yang2017dagan,schlemper2017deep,qin2018convolutional,sun2016deep,mardani2018deep,lee2017deep,han2018deep,wang20171d,quan2016compressed,seitzer2018adversarial,zhang2018multi,duan2019vs,schlemper2018stochastic,quan2018compressed,hammernik2018learning}. For example, a conditional Generative Adversarial Networks-based model (DAGAN) was proposed to achieve fast CS-MRI \cite{yang2017dagan}, but still, this end-to-end training neglected the correlation between adjacent 2D slices. Thus, although DAGAN can achieve fast MRI reconstruction, it may lose image quality without using a priori information. For another example, DC-CNN \cite{schlemper2017deep} applied cascades of convolutional neural networks with a residual connection for CS-MRI. Besides, DC-CNN also used a data consistency (DC) step to ensure that the output of each cascade was consistent with the original \textit{k}-space information. However, DC-CNN approach was not able to effectively utilise the full temporal domain information. In contrast, a convolutional recurrent neural network (CRNN) method was proposed to incorporate a bidirectional convolutional recurrent unit for a faster and more stable reconstruction \cite{qin2018convolutional}. However, such an approach was not able to effectively exploit the \textit{k}-space information from individual images.
\begin{figure}[t!]
\includegraphics[width=\textwidth]{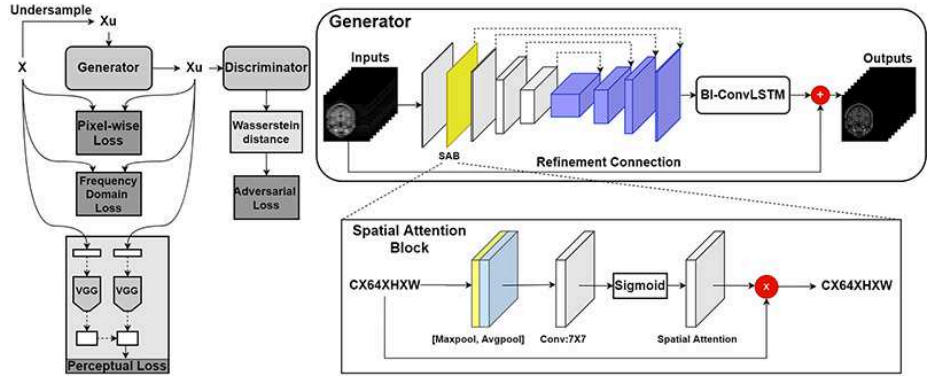}
\caption{The overall network architecture of our DAWGAN framework. Left: the workflow of our proposed model. Right: the details of our proposed generator with Bi-ConvLSTM and flow diagram of the proposed spatial attention block (SAB).} \label{fig1}
\end{figure}

In this study, we propose a GAN based architecture that works on continuous sequential data for CS-MRI. This intuitively mimics the way reporting clinicians scrutinise the 3D data by scrolling up and down to fully sense the information above and below the current 2D slice. Our method can not only overcome the shortcomings of slow reconstruction but can also maintain higher reconstructed image quality by combining the characteristics in time and frequency domains. To the best of our knowledge, this is the first study to combine Recurrent Neural Networks (RNN) with GAN in the field of MRI reconstruction. In particular, we design a novel generator with bidirectional convolutional long short-term memory (Bi-ConvLSTM) that can encode the a priori frequency and time-domain information. Besides, another significant contribution of our work is that we propose a spatial attention-based model that the attention unit in our model can distinguish between significant and non-significant features in terms of the MRI reconstruction task. In addition, we utilise WGAN with gradient penalty (WGAN-GP) as a critic function, which can significantly improve the stability of GAN. We also couple the adversarial loss with pixel-wise mean square error (MSE) and the perceptual loss ~\cite{johnson2016perceptual} to achieve better reconstruction details with superior perceptual image quality.

\section{Method}
\subsection{Problem formulation of CS-MRI}

\subsubsection{Deep Learning-based CS-MRI Reconstruction.}
Let $\mathbf{x} \in \mathbb{C}^{D}$ represents the slice of 2D images to be reconstructed, which consists of $\sqrt{N} \times \sqrt{N}$ pixels for one image, and let $\mathbf{y}$ denotes the undersampled measurements in \textit{k}-space. For deep learning-based methods, previous studies such as \cite{wang2016accelerating} and \cite{schlemper2017deep} incorporated a CNN into CS-MRI reconstruction, transformed the unconstrained optimisation problem into:
\begin{equation}
\min _{\mathbf{x}} \lambda\left\|\mathbf{y}-\mathbf{F}_{u} \mathbf{x}\right\|_{2}^{2}+\mathcal{R}(\mathbf{x})+\zeta\left\|\mathbf{x}-f_{\mathrm{cnn}}\left(\mathbf{X}_{u} | \hat{\theta}\right)\right\|_{2}^{2},
\end{equation}
in which $f_{\mathrm{cnn}}$ denotes the forward propagation of data through the CNN paramet-rised by $\theta$, and $\zeta$ is a regularisation parameter. $\mathbf{X}_{u}$ is the reconstruction from the zero-filled undersampled \textit{k}-space measurements. 

In the reconstruction network selection, many previous studies, e.g., \cite{yang2017dagan,lee2017deep,han2018deep,wang20171d}, relied on an encoder-decoder structure. Nevertheless, our preliminary experiments indicated that these single structures performed poorly in the PSNR. Moreover, there are also methods \cite{schlemper2017deep,qin2018convolutional} that developed for the dynamic MR reconstruction, but they did not perform well at higher \textit{k}-space undersampling.

\subsection{DAWGAN for CS-MRI}
In this study, we propose a Deep Attentive Wasserstein Generative Adversarial Networks (DAWGAN) method to reconstruct MRI images from highly undersampled data with continuous sequential data. It contains three key components: a Bi-ConvLSTM block, a spatial attention block (SAB) and a WGAN-GP as the critic function. The workflow of our DAWGAN is summarised in Figure 1.



\subsubsection{Image Domain Feature Extraction via a Sequential Learning.}
To achieve more aggressive undersampling, one way is to encode the a priori frequency and time-domain information in sequential data, e.g., 2D MRI slices of a 3D volumetric data. We assumed $\mathbf{X}$ as the feature representation of our 2D sequential MRI data slices throughout the 3D volume. Here $\mathbf{X}_{l}^{(i)}$ denoted the representation at slice $\mathbf{l}$ and iteration $\mathbf{i}$. We needed to take into account $\mathbf{X}_{l-1}^{(i)}$ and $\mathbf{X}_{l+1}^{(i)}$ in the reconstruction process to provide information for $\mathbf{X}_{l}^{(i)}$. To that end, we proposed a Bi-ConvLSTM subnetwork to exploit both temporal and iteration dependencies jointly. The Bi-ConvLSTM subnetwork can be formulated as:
\begin{equation}
\begin{array}{l}
{\overrightarrow{\mathbf{X}}_{l, t}^{(i)}=\sigma\left(\mathbf{W}_{l} * \mathbf{X}_{l-1, t}^{(i)}+\mathbf{W}_{t} * \overrightarrow{\mathbf{X}}_{l, t-1}^{i}+\mathbf{W}_{i} * \mathbf{X}_{l, t}^{(i-1)}+\overrightarrow{\mathbf{B}}_{l}\right)} \\
{\overleftarrow{\mathbf{X}}_{l, t}^{(i)}=\sigma\left(\mathbf{W}_{l} * \mathbf{X}_{l-1, t}^{(i)}+\mathbf{W}_{t} * \overleftarrow{\mathbf{X}}_{l, t+1}^{(i)}+\mathbf{W}_{i} * \mathbf{X}_{l, t}^{(i-1)}+\overleftarrow{\mathbf{B}}_{l}\right)} \\
{\mathbf{X}_{l, t}^{(i)}=\overrightarrow{\mathbf{X}}_{l, t}^{(i)}+\overleftarrow{\mathbf{X}}_{l, t}^{(i)}}
\end{array}
\end{equation}
where $\overrightarrow{\mathbf{X}}_{l, t}^{(i)}$ denoted the forward direction and $\overleftarrow{\mathbf{X}}_{l, t}^{(i)}$ denoted the backward direction. Through Bi-ConvLSTM layer, our model can learn the differences and correlations of successive MRI data slices. The output of the Bi-ConvLSTM layer then took a refinement connection to prevent data shifting. 

\subsubsection{Spatial Attention Block (SAB).}
The main aim of the designed SAB was to increase representation power by using attention mechanism: focusing on important features and suppressing unnecessary ones. Details about the SAB are shown in Figure \ref{fig1}. Inspired by \cite{woo2018cbam}, we set the SAB after the first convolution block, which was also propagated to the up-sampling layers with the skip-connections. We conducted average-pooling and max-pooling operations on the feature map obtained from the upper layer to generate an efficient feature descriptor. Then we utilised a convolution layer to generate a feature map that could encode where to emphasize or suppress. The SAB took all the features extracted by the upper layer to calculate the attention map.

We assumed that the 2D maps generated by pooling operations were $\mathbf{F}_{\text {avg }} \in \mathbb{R}^{1 \times H \times W}$ and $\mathbf{F}_{\text {max }} \in \mathbb{R}^{1 \times H \times W}$. Each denoted average-pooled features and max-pooled features across the feature map. The two maps were then stacked and convolved by a standard convolution layer to produce the 2D spatial attention map. Hence, our spatial attention map was computed as
\begin{equation}
\begin{aligned}
\mathbf{M}_{\mathbf{s}}(\mathbf{F}) =\sigma\left(f^{7 \times 7}\left(\left[\mathbf{F}_{\text {avg }} ; \mathbf{F}_{\mathrm{max}}\right]\right)\right)
\end{aligned}
\end{equation}
where $f^{7 \times 7}$ represented the convolution operation with the filter size of 7$\times$7 and $\sigma$ denoted the sigmoid function according to \cite{woo2018cbam}. The spatial attention calculated the feature correlation across the channel domain to find the cardinal features across the entire spatial domain.

\subsubsection{Loss Function and Training.}
Our loss function consisted of content loss and adversarial loss. The content loss function was basically made up of three parts, i.e., a pixel-wise image domain mean square error (MSE) loss, a frequency domain MSE loss and a perceptual VGG loss. The whole loss function could be formulated as
\begin{equation}
\mathcal{L}_{\mathrm{TOTAL}}=\alpha \mathcal{L}_{\mathrm{iMSE}}+\beta \mathcal{L}_{\mathrm{fMSE}}+\gamma \mathcal{L}_{\mathrm{VGG}}+\mathcal{L}_{\mathrm{GEN}}
\end{equation}
where $\alpha, \beta, \gamma$ represented the hyper-parameters.

\begin{figure}[t!]
\includegraphics[width=\textwidth]{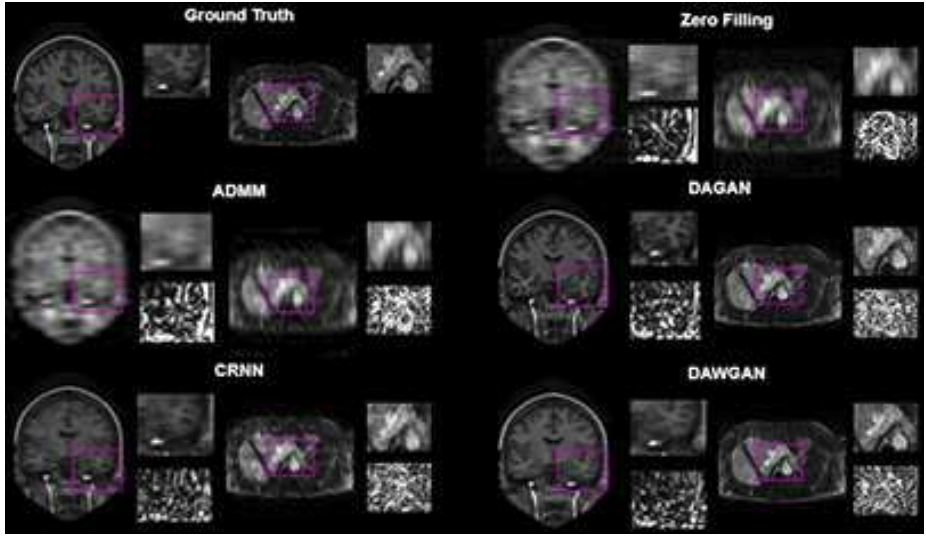}
\caption{Qualitative visualisation using 10\% of the \textit{k}-space data. For each subfigure, left: brain MRI data; right: cardiac MRI data.} \label{fig2}
\end{figure}
Most of the GAN based CS-MRI studies used vanilla GAN objective~\cite{goodfellow2014generative}, which applied the Kullback-Leibler (KL) divergence, as the adversarial loss function. However, during the training of the generator, when the generator deviated from the optimal solution, the parameters of the generator might not be updated continuously, which could then lead to complicated training process and \textit{model collapse} ~\cite{arjovsky2017wasserstein}. In this study, we introduced WGAN-GP~\cite{gulrajani2017improved} as an alternative strategy of using Wasserstein distance to displace the KL divergence for solving the potential complications in training the GAN. WGAN-GP also introduced the \textit{gradient penalty} to better solve the common gradient vanishing problem. We used a loss that was calculated as the following
\begin{equation}
\mathcal{L}_{\mathrm{GEN}}=-E_{x \sim p_{g}}\left[f_{\mathrm{model}}(x)\right].
\end{equation}

In order to improve the perceptual quality, we also incorporated the content loss with three different combinations of the loss functions:
\begin{equation}\begin{aligned}
\min _{\theta_{G}} \mathcal{L}_{\mathrm{iMSE}}\left(\theta_{G}\right) &=\frac{1}{2}\left\|\mathrm{x}_{t}-\hat{\mathrm{x}}_{u}\right\|_{2}^{2} \\
\min _{\theta_{G}} \mathcal{L}_{\mathrm{fMSE}}\left(\theta_{G}\right) &=\frac{1}{2}\left\|\mathrm{y}_{t}-\hat{\mathrm{y}}_{u}\right\|_{2}^{2} \\
\min _{\theta_{G}} \mathcal{L}_{\mathrm{VGG}}\left(\theta_{G}\right) &=\frac{1}{2}\left\|f_{\mathrm{Vgg}}\left(\mathrm{x}_{t}\right)-f_{\mathrm{vgg}}\left(\hat{x}_{u}\right)\right\|_{2}^{2}.
\end{aligned}\end{equation}
We used normalised MSE (NMSE) as the optimisation cost function. However, the use of NMSE as content loss alone might lead to perceptually uneven reconstruction, resulting in a lack of coherent image details. Therefore, to consider the perceptual similarity of images, we also added NMSE of the frequency domain data and VGG loss ($\mathcal{L}_{\mathrm{VGG}}$) as additional constraints.

\section{Experiments and Results}
\begin{table}[t!]
\begin{center} 
\caption{Comparison study results using different CS-MRI methods.}
\scriptsize
\begin{tabular}{cllllll}
\hline
\multicolumn{7}{c}{\textbf{Brain MRI Data}} \\ \hline
\multirow{2}{*}{\textbf{Methods}} &
  \multicolumn{2}{c}{\textbf{10\%}} &
  \multicolumn{2}{c}{\textbf{30\%}} &
  \multicolumn{2}{c}{\textbf{50\%}} \\ \cline{2-7} 
 &
  \multicolumn{1}{c}{\textbf{PSNR}} &
  \multicolumn{1}{c}{\textbf{MOS}} &
  \multicolumn{1}{c}{\textbf{PSNR}} &
  \multicolumn{1}{c}{\textbf{MOS}} &
  \multicolumn{1}{c}{\textbf{PSNR}} &
  \multicolumn{1}{c}{\textbf{MOS}} \\ \hline
\multicolumn{1}{c|}{\textbf{Zero-filling}} &
  28.16(3.33) &
  \multicolumn{1}{l|}{1.02(0.13)} &
  34.83(2.78) &
  \multicolumn{1}{l|}{1.12(0.21)} &
  39.36(2.61) &
  1.09(0.34) \\
\multicolumn{1}{c|}{\textbf{ADMM}} &
  28.20(3.36) &
  \multicolumn{1}{l|}{1.21(0.35)} &
  35.21(4.03) &
  \multicolumn{1}{l|}{1.22(0.31)} &
  39.99(4.08) &
  1.28(0.37) \\
\multicolumn{1}{c|}{\textbf{DAGAN}} &
  33.25(4.10) &
  \multicolumn{1}{l|}{2.52(0.88)} &
  38.12(3.56) &
  \multicolumn{1}{l|}{3.08(0.68)} &
  45.41(4.13) &
  3.27(0.73) \\
\multicolumn{1}{c|}{\textbf{CRNN}} &
  33.57(3.16) &
  \multicolumn{1}{l|}{2.78(0.62)} &
  38.26(3.86) &
  \multicolumn{1}{l|}{2.98(0.43)} &
  46.10(2.29) &
  3.58(0.61) \\
\multicolumn{1}{c|}{\textbf{DAWGAN}} &
  \textbf{34.31(3.01)} &
  \multicolumn{1}{l|}{\textbf{3.01(0.69)}} &
  \textbf{40.74(3.57)} &
  \multicolumn{1}{l|}{\textbf{3.23(0.69)}} &
  \textbf{46.43(2.19)} &
  \textbf{3.98(0.72)} \\ \hline
\multicolumn{7}{c}{\textbf{Cardiac MRI Data}} \\ \hline
\multicolumn{1}{c|}{\textbf{Zero-filling}} &
  27.08(0.84) &
  \multicolumn{1}{l|}{1.02(0.13)} &
  31.49(0.88) &
  \multicolumn{1}{l|}{1.12(0.21)} &
  35.13(0.92) &
  1.09(0.34) \\
\multicolumn{1}{c|}{\textbf{ADMM}} &
  27.20(1.64) &
  \multicolumn{1}{l|}{1.25(0.28)} &
  31.88(1.72) &
  \multicolumn{1}{l|}{1.38(0.21)} &
  35.54(1.73) &
  1.98(0.45) \\
\multicolumn{1}{c|}{\textbf{DAGAN}} &
  29.35(1.33) &
  \multicolumn{1}{l|}{2.21(0.54)} &
  33.85(1.62) &
  \multicolumn{1}{l|}{2.49(0.41)} &
  37.86(1.22) &
  2.81(0.53) \\
\multicolumn{1}{c|}{\textbf{CRNN}} &
  29.62(2.15) &
  \multicolumn{1}{l|}{2.68(0.61)} &
  34.29(2.33) &
  \multicolumn{1}{l|}{2.83(0.71)} &
  38.12(2.29) &
  2.96(0.67) \\
\multicolumn{1}{c|}{\textbf{DAWGAN}} &
  \textbf{31.06(1.71)} &
  \multicolumn{1}{l|}{\textbf{2.92(0.72)}} &
  \textbf{35.97(1.77)} &
  \multicolumn{1}{l|}{\textbf{3.06(0.59)}} &
  \textbf{39.66(1.79)} &
  \textbf{3.42(0.56)} \\ \hline
\end{tabular}\label{table1}
\end{center}
\end{table}
\subsection{Experiments}
\begin{figure}[b!]
\includegraphics[width=\textwidth,height=3.5cm]{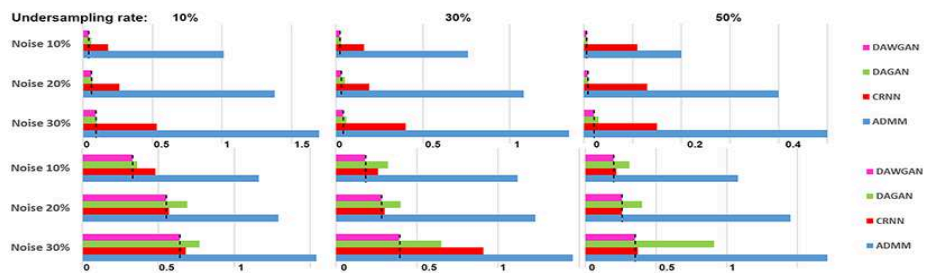}
\caption{Estimated residual noise from the reconstructed images with respect to different noise levels. Upper panel: brain MRI data; Lower panel: cardiac MRI data.} \label{fig3}
\end{figure}
\subsubsection{Datasets.}
Our experiments were performed on two datasets (1) Brain MRI dataset: We trained and tested our model using a MICCAI 2013 grand challenge dataset. In total, we included 726 3D data for our study. We randomly used 503 data for training, 173 for validation and 50 for testing. Each 2D slice had a shape of 256$\times$256, and we normalised the intensities into a range of [-1 1]. (2) Cardiac MRI dataset: A population of 100 3D LGE MRI patient data, which were made available through the 2018 Atrial Segmentation Challenge, were used in this work. The scanner used for this clinical study was a whole-body MRI scanner, within an image acquisition resolution of 0.625mm$^3$. The studied data were randomly divided into 80 for training, 10 for validation and 10 for testing. Similarly, we normalised each slice into a range of [-1 1].


\subsubsection{Experiments Setup.}
For all the input data, we applied data augmentation on the input 2D image slices. Besides, we used raw \textit{k}-space data with different undersampling ratios to simulate the corresponding acceleration factors. In particular, $10\%, 30\%$ and $50\%$ retained raw \textit{k}-space data were simulated representing $10\times, 3.3\times, 2\times$ accelerations assuming that the preparation time of MRI scanning is insignificant. All our comparison studies were carried out using different CS-MRI reconstruction algorithms using these three levels of undersampling ratios. Our studies were mainly divided into three experiments: First, We compared the performance of our method with that of other SOTA at three different acceleration factors. In addition to the traditional metrics of PSNR, we also introduced the mean opinion scores (MOS) to take human perception into account, which was the results of domain experts evaluating the reconstructions and averaging their perceptual quality. Then, at different acceleration factors, we tested the noise reduction effect of all models at different noise level to prove that our model could significantly suppress the residual noise. To test the noise tolerance of different CS-MRI methods, we added white Gaussian noises to the \textit{k}-space data before applying the undersampling. Inspired by \cite{liu2013single}, we conducted a noise level estimation for all the reconstruction results. Finally, we tested the effectiveness of various network configurations of our proposed framework. In the final quantification, we used PSNR, SSIM and NMSE as the evaluation metrics.


\subsection{Results}
\begin{figure}[b!]
\includegraphics[width=\textwidth,height=4cm]{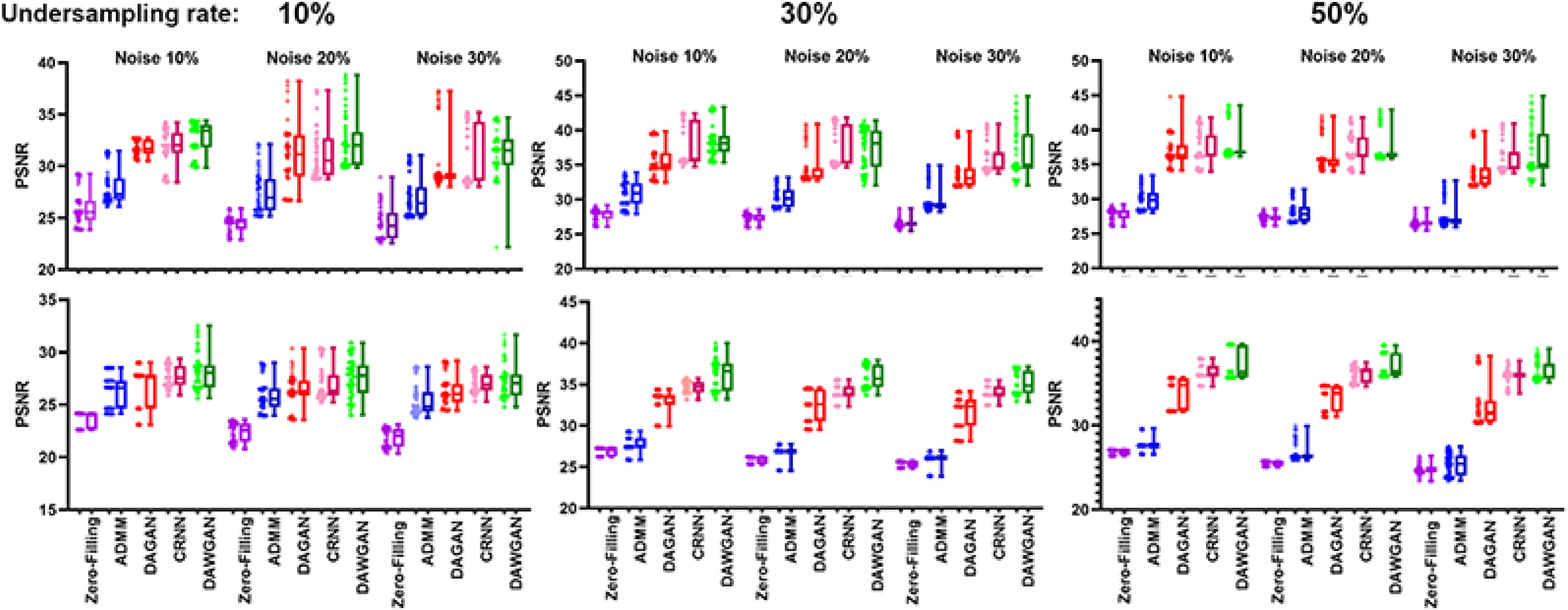}
\caption{PSNR with respect to different noise levels at various undersampling ratios, i.e., 10\%, 30\% and 50\%, respectively. The upper panel shows the results of the brain MRI dataset, and the lower panel shows the results of the cardiac MRI dataset.} \label{fig4}
\end{figure}
\subsubsection{Image Quality Comparison.}
Our method has demonstrated the best performance by comparing with four SATO methods (on both brain and cardiac MRI datasets). Table \ref{table1} shows that the results of our proposed DAWGAN performed best in PSNR and MOS. At the $10\times$ and $3.3\times$ acceleration factors, the PSNR and MOS achieved by our DAWGAN were significantly higher than the other methods. In addition, Figure \ref{fig2} shows that DAWGAN produced less noise-induced artefacts in all the simulation studies, while the other methods had more noise-induced artefacts. Although CRNN and DAGAN could also suppress some artefacts, the reconstructions of the brain area were less detailed than those DAWGAN reconstructed. Moreover, ADMM and zero-filling could not effectively inhibit remaining aliasing artefacts.
\subsubsection{Noise Suppression Comparison.}
In terms of reconstruction details, we demonstrated that DAWGAN could effectively reduce residual noise in the reconstructed images. As shown in Figure \ref{fig3}, DAWGAN suppressed the noise effectively at different noise levels. Figure \ref{fig4} shows the PSNR results with respect to different noise levels and various undersampling patterns. Our proposed DAWGAN also demonstrated considerable noise tolerance at different noise levels, and the mean value of the PSNR was higher than other methods. 
\subsubsection{Ablation Studies.}
The performance of our framework with various network components was shown by our ablation studies. The sub-models we compared were WGAN-GP+RNN, WGAN-GP+Attention and Attention+RNN. Table \ref{table2} shows that the DAWGAN full model was superior to other sub-model variations using all three metrics, which indicated that the current configurations in our proposed network architecture are effective.
\begin{table}[t!]
\caption{Ablation study results of our framework with various network configurations.}
\resizebox{\textwidth}{!}{\begin{tabular}{llllllllll}
\hline
\multicolumn{10}{c}{\textbf{Brain MRI Data}} \\ \hline
\multicolumn{1}{c}{\multirow{2}{*}{\textbf{Comparison Study}}} &
  \multicolumn{3}{c}{\textbf{10\%}} &
  \multicolumn{3}{c}{\textbf{30\%}} &
  \multicolumn{3}{c}{\textbf{50\%}} \\ \cline{2-10} 
\multicolumn{1}{c}{} &
  \multicolumn{1}{c}{\textbf{PSNR}} &
  \multicolumn{1}{c}{\textbf{NMSE}} &
  \multicolumn{1}{c}{\textbf{SSIM}} &
  \multicolumn{1}{c}{\textbf{PSNR}} &
  \multicolumn{1}{c}{\textbf{NMSE}} &
  \multicolumn{1}{c}{\textbf{SSIM}} &
  \multicolumn{1}{c}{\textbf{PSNR}} &
  \multicolumn{1}{c}{\textbf{NMSE}} &
  \multicolumn{1}{c}{\textbf{SSIM}} \\ \hline
\multicolumn{1}{l|}{WGAN-GP+RNN} &
  34.11(3.91) &
  0.16(0.03) &
  \multicolumn{1}{l|}{0.96(0.02)} &
  40.95(4.13) &
  0.07(0.02) &
  \multicolumn{1}{l|}{0.98(0.01)} &
  46.08(4.37) &
  0.03(0.01) &
  0.99(0.00) \\
\multicolumn{1}{l|}{WGAN-GP+Attention} &
  34.22(2.01) &
  0.17(0.03) &
  \multicolumn{1}{l|}{0.96(0.02)} &
  39.90(3.57) &
  0.08(0.02) &
  \multicolumn{1}{l|}{0.98(0.01)} &
  46.29(3.81) &
  0.03(0.01) &
  0.99(0.00) \\
\multicolumn{1}{l|}{Attention+RNN} &
  33.71(3.30) &
  0.16(0.03) &
  \multicolumn{1}{l|}{0.96(0.02)} &
  40.74(3.57) &
  0.08(0.02) &
  \multicolumn{1}{l|}{0.98(0.01)} &
  46.29(3.81) &
  0.03(0.01) &
  0.99(0.00) \\
\multicolumn{1}{l|}{DAWGAN} &
  \textbf{34.31(3.01)} &
  \textbf{0.16(0.03)} &
  \multicolumn{1}{l|}{\textbf{0.96(0.02)}} &
  \textbf{40.74(3.57)} &
  \textbf{0.07(0.02)} &
  \multicolumn{1}{l|}{\textbf{0.98(0.01)}} &
  \textbf{46.43(2.19)} &
  \textbf{0.03(0.01)} &
  \textbf{0.99(0.00)} \\ \hline
\multicolumn{10}{c}{\textbf{Cardiac MRI Data}} \\ \hline
\multicolumn{1}{l|}{WGAN-GP+RNN} &
  30.62(2.33) &
  0.24(0.03) &
  \multicolumn{1}{l|}{0.89(0.02)} &
  34.51(2.40) &
  0.15(0.02) &
  \multicolumn{1}{l|}{0.94(0.01)} &
  38.44(2.51) &
  0.09(0.01) &
  0.97(0.00) \\
\multicolumn{1}{l|}{WGAN-GP+Attention} &
  30.68(2.61) &
  0.23(0.03) &
  \multicolumn{1}{l|}{0.89(0.02)} &
  34.81(2.13) &
  0.14(0.02) &
  \multicolumn{1}{l|}{0.94(0.01)} &
  37.87(2.54) &
  0.10(0.01) &
  0.96(0.00) \\
\multicolumn{1}{l|}{Attention+RNN} &
  30.51(2.85) &
  0.23(0.03) &
  \multicolumn{1}{l|}{0.87(0.02)} &
  34.35(2.70) &
  0.15(0.02) &
  \multicolumn{1}{l|}{0.93(0.01)} &
  37.67(2.10) &
  0.10(0.01) &
  0.96(0.00) \\
\multicolumn{1}{l|}{DAWGAN} &
  \textbf{31.06(1.71)} &
  \textbf{0.23(0.02)} &
  \multicolumn{1}{l|}{\textbf{0.90(0.02)}} &
  \textbf{35.97(1.77)} &
  \textbf{0.13(0.01)} &
  \multicolumn{1}{l|}{\textbf{0.96(0.01)}} &
  \textbf{39.66(1.79)} &
  \textbf{0.09(0.01)} &
  \textbf{0.97(0.00)} \\ \hline
\end{tabular}}\label{table2}
\end{table}
\section{Conclusion}
In this paper, we proposed the DAWGAN to reconstruct MRI images from highly undersampled \textit{k}-space data. Our DAWGAN employed WGAN-GP to improve the stability of vanilla GAN. The incorporated Bi-ConvLSTM block can make full use of the relationships among successive MRI slices to improve the reconstruction results. In addition, the proposed SAB can distinguish between significant and non-significant features for our MRI reconstruction task. Our ablation studies have demonstrated the effectiveness of the key components of our framework. The comprehensive comparison studies on both brain and cardiac MRI datasets have corroborated that our method can not only achieve better reconstruction results but can also effectively reduce residual noise generated during the reconstruction process.
\bibliographystyle{splncs04}

\begin{thebibliography}{8}




\bibitem{donoho2006compressed}
Donoho, D.L., et~al.: Compressed sensing. IEEE Transactions on Information
  Theory  \textbf{52}(4),  1289--1306 (2006)

\bibitem{hollingsworth2015reducing}
Hollingsworth, K.G.: Reducing acquisition time in clinical mri by data
  undersampling and compressed sensing reconstruction. Physics in Medicine \&
  Biology  \textbf{60}(21), ~R297 (2015)

\bibitem{Ma2008An}
Ma, S., Yin, W., Zhang, Y., Chakraborty, A.: An efficient algorithm for
  compressed mr imaging using total variation and wavelets. In: 2008 IEEE
  Conference on Computer Vision and Pattern Recognition (2008)

\bibitem{wang2016accelerating}
Wang, S., Su, Z., Ying, L., Peng, X., Zhu, S., Liang, F., Feng, D., Liang, D.:
  Accelerating magnetic resonance imaging via deep learning. In: 2016 IEEE 13th
  International Symposium on Biomedical Imaging. pp. 514--517. IEEE
  (2016)

\bibitem{yang2017dagan}
Yang, G., Yu, S., Dong, H., Slabaugh, G., Dragotti, P.L., Ye, X., Liu, F.,
  Arridge, S., Keegan, J., Guo, Y., et~al.: Dagan: deep de-aliasing generative
  adversarial networks for fast compressed sensing mri reconstruction. IEEE
  Transactions on Medical Imaging  \textbf{37}(6),  1310--1321 (2017)

\bibitem{schlemper2017deep}
Schlemper, J., Caballero, J., Hajnal, J.V., Price, A.N., Rueckert, D.: A deep
  cascade of convolutional neural networks for dynamic mr image reconstruction.
  IEEE Transactions on Medical Imaging  \textbf{37}(2),  491--503 (2017)

\bibitem{qin2018convolutional}
Qin, C., Schlemper, J., Caballero, J., Price, A.N., Hajnal, J.V., Rueckert, D.:
  Convolutional recurrent neural networks for dynamic mr image reconstruction.
  IEEE Transactions on Medical Imaging  \textbf{38}(1),  280--290 (2018)

\bibitem{sun2016deep}
Sun, J., Li, H., Xu, Z., et~al.: Deep admm-net for compressive sensing mri. In:
  Advances in Neural Information Processing Systems. pp. 10--18 (2016)

\bibitem{mardani2018deep}
Mardani, M., Gong, E., Cheng, J.Y., Vasanawala, S.S., Zaharchuk, G., Xing, L.,
  Pauly, J.M.: Deep generative adversarial neural networks for compressive
  sensing mri. IEEE Transactions on Medical Imaging  \textbf{38}(1),  167--179
  (2018)

\bibitem{lee2017deep}
Lee, D., Yoo, J., Ye, J.C.: Deep residual learning for compressed sensing mri.
  In: 2017 IEEE 14th International Symposium on Biomedical Imaging.
  pp. 15--18. IEEE (2017)

\bibitem{johnson2016perceptual}
Johnson, J., Alahi, A., Fei-Fei, L.: Perceptual losses for real-time style
  transfer and super-resolution. In: European Conference on Computer Vision.
  pp. 694--711. Springer (2016)

\bibitem{woo2018cbam}
Woo, S., Park, J., Lee, J.Y., So~Kweon, I.: Cbam: Convolutional block attention
  module. In: Proceedings of the European Conference on Computer Vision (ECCV).
  pp. 3--19 (2018)

\bibitem{goodfellow2014generative}
Goodfellow, I., Pouget-Abadie, J., Mirza, M., Xu, B., Warde-Farley, D., Ozair,
  S., Courville, A., Bengio, Y.: Generative adversarial nets. In: Advances in
  Neural Information Processing Systems. pp. 2672--2680 (2014)

\bibitem{arjovsky2017wasserstein}
Arjovsky, M., Chintala, S., Bottou, L.: Wasserstein gan. arXiv preprint
  arXiv:1701.07875  (2017)

\bibitem{gulrajani2017improved}
Gulrajani, I., Ahmed, F., Arjovsky, M., Dumoulin, V., Courville, A.C.: Improved
  training of wasserstein gans. In: Advances in Neural Information Processing
  Systems. pp. 5767--5777 (2017)

\bibitem{liu2013single}
Liu, X., Tanaka, M., Okutomi, M.: Single-image noise level estimation for blind
  denoising. IEEE Transactions on Image Processing  \textbf{22}(12),
  5226--5237 (2013)

\bibitem{han2018deep}
Han, Y., Yoo, J., Kim, H.H., Shin, H.J., Sung, K., Ye, J.C.: Deep learning with
  domain adaptation for accelerated projection-reconstruction mr. Magnetic
  Resonance in Medicine  \textbf{80}(3),  1189--1205 (2018)
  
\bibitem{wang20171d}
Wang, S., Huang, N., Zhao, T., Yang, Y., Ying, L., Liang, D.: 1d partial
  fourier parallel mr imaging with deep convolutional neural network. In:
  Proceedings of the 25st Annual Meeting of ISMRM, Honolulu, HI, USA. vol.~1,
  p.~2 (2017)

\bibitem{quan2016compressed}
Quan, T.M., Jeong, W.K.: Compressed sensing dynamic mri reconstruction using
  gpu-accelerated 3d convolutional sparse coding. In: International Conference
  on Medical Image Computing and Computer-Assisted Intervention. pp. 484--492.
  Springer (2016)
  
\bibitem{seitzer2018adversarial}
Seitzer, M., Yang, G., Schlemper, J., Oktay, O., W{\"u}rfl, T., Christlein, V.,
  Wong, T., Mohiaddin, R., Firmin, D., Keegan, J., et~al.: Adversarial and
  perceptual refinement for compressed sensing mri reconstruction. In:
  International Conference on Medical Image Computing and Computer-Assisted
  Intervention. pp. 232--240. Springer (2018)
  
\bibitem{zhang2018multi}
Zhang, P., Wang, F., Xu, W., Li, Y.: Multi-channel generative adversarial
  network for parallel magnetic resonance image reconstruction in k-space. In:
  International Conference on Medical Image Computing and Computer-Assisted
  Intervention. pp. 180--188. Springer (2018)
  
\bibitem{duan2019vs}
Duan, J., Schlemper, J., Qin, C., Ouyang, C., Bai, W., Biffi, C., Bello, G.,
  Statton, B., O’Regan, D.P., Rueckert, D.: Vs-net: Variable splitting
  network for accelerated parallel mri reconstruction. In: International
  Conference on Medical Image Computing and Computer-Assisted Intervention. pp.
  713--722. Springer (2019) 
  
\bibitem{schlemper2018stochastic}
Schlemper, J., Yang, G., Ferreira, P., Scott, A., McGill, L.A., Khalique, Z.,
  Gorodezky, M., Roehl, M., Keegan, J., Pennell, D., et~al.: Stochastic deep
  compressive sensing for the reconstruction of diffusion tensor cardiac mri.
  In: International Conference on Medical Image Computing and Computer-Assisted
  Intervention. pp. 295--303. Springer (2018)
  
\bibitem{quan2018compressed}
Quan, T.M., Nguyen-Duc, T., Jeong, W.K.: Compressed sensing mri reconstruction
  using a generative adversarial network with a cyclic loss. IEEE Transactions
  on Medical Imaging  \textbf{37}(6),  1488--1497 (2018)
  
  
\bibitem{hammernik2018learning}
Hammernik, K., Klatzer, T., Kobler, E., Recht, M.P., Sodickson, D.K., Pock, T.,
  Knoll, F.: Learning a variational network for reconstruction of accelerated
  mri data. Magnetic Resonance in Medicine  \textbf{79}(6),  3055--3071 (2018)
\end{thebibliography}
%

\end{document}